\documentclass[prb,twocolumn]{revtex4}
\usepackage{graphicx}
\usepackage{dcolumn}
\usepackage{bm}
\begin{document}
\draft
\def \beq{\begin{equation}}
\def \eeq{\end{equation}}
\def \beqarr{\begin{eqnarray}}
\def \eeqarr{\end{eqnarray}}

%\twocolumn[\hsize\textwidth\columnwidth\hsize\csname @twocolumnfalse\endcsname

\title{
Vortex Lattice Structure of Fulde-Ferrell-Larkin-Ovchinnikov
Superconductors
}

\author{Kun Yang}

\address{
National High Magnetic Field Laboratory and 
Department of Physics,
Florida State University, Tallahassee, Florida 32306, USA
}

\author{A. H. MacDonald}

\address{
Department of Physics, University of Texas, Austin, Texas 78712, USA
}

\date{\today}

\begin{abstract}
In superconductors with singlet pairing, the
inhomogeneous Fulde-Ferrell-Larkin-Ovchinnikov (FFLO) 
state is expected to be stabilized by a large Zeeman splitting.
We develop an efficient method to evaluate the Landau-Ginzburg free 
energies of FFLO-state vortex lattices and use it to simplify the considerations 
that determine the optimal vortex configuration at different points in the phase
diagram.  
%Allan
We demonstrate that the order parameter spatial profile is completely determined, up 
to a uniform translation, by its Landau level index $n$ and the vortex Lattice structure and 
derive an explicit expression for the order parameter spatial profile that can be used 
to determine $n$ from experimental data. 
%One can nevertheless determine the Landau level index from the
%number of antivortices per magnetic flux quantum, {\em without} the 
%knowledge of
%detailed lattice structure. This allows for an estimate of
%the order parameter momentum. 
\end{abstract}


\maketitle

\section{introduction}
Forty years ago Fulde and Ferrell\cite{ff},
and Larkin and Ovchinnikov\cite{lo} proposed that a superconducting state with
an inhomogeneous order parameter would become stable when a singlet superconductor
is disturbed by Zeeman splitting between electrons with opposite spin orientations.
% that is comparable to the quasiparticle energy gap.  
The Zeeman splitting can be
due to either an external magnetic field or an internal exchange field.
In spite of ongoing theoretical interest\cite{review} in the
Fulde-Ferrell-Larkin-Ovchinnikov(FFLO) state, evidence for
its occurrence has been sparse until recently.  The situation has been changed
by experimental results suggestive of the FFLO state 
in organic, heavy-fermion, and possibly other unconventional supercondcutors
\cite{gloos,modler,tachiki,geg,singleton,radovan,bianchi,martin}.
Recent experimental results on the heavy fermion compound
CeCoIn$_5$, a quasi-two-dimensional d-wave superconductor,
are particularly compelling\cite{radovan,bianchi,martin}.
Independently, speculations that the FFLO state may also be realized in high
density quark matter have caught the attention of the 
particle physics community\cite{review}, adding to the flurry 
of interest in this uncommon superconducting state. 

To establish the presence of the FFLO state unambiguously one needs to
perform a {\em phase sensitive} experiment that detects order parameter 
oscillations and measures their wavelength.  One possible phase sensitive probe
based on the Josephson effect between a FFLO superconductor and a 
BCS superconductor\cite{ya}
was proposed some time ago by Yang and Agterberg.  Although
conceptually straightforward this experiment is yet to be carried out,
possibly due to technical difficulties in its implementation. More recently,
there has been interest in the vortex lattice structure (VLS)
of an FFLO state subjected to an orbital magnetic field, and particularly 
in the possibility of using the character of the 
VLS to infer the presence of an underlying  
FFLO superconductor\cite{sr,krs,hb,hbbm,klein}.  The key idea\cite{bul} 
is that, depending on the
interplay between the orbital and Zeeman effects of the magnetic field,
the order parameter of a FFLO state near its upper 
critical field can correspond to a high
Landau level (LL) index Cooper pair wave function. 
Recent work\cite{sr,krs,hb,hbbm,klein} on the FFLO VLS 
in specific situations has demonstrated that these high LL index VLS's 
can be very different from the triangular lattice Abrikosov VLS favored by lowest Landau level (LL) Cooper pairs.
The case of FFLO states in an orbital field is the only example known to us
in which bosons (Cooper pairs in the present case) are placed in high LLs; the absence of 
a fermionic Pauli exclusion principle implies that one cannot force bosons to occupy a high LL simply 
by increasing the particle density.  Electronic states in 
high LLs have recently been found to exhibit physics\cite{electron} that can be quite different 
from that found in the lowest LL, suggesting that surprises might also occur in 
high LL bosonic physics.

In this paper we apply techniques developed in studies of electronic states 
in a strong magnetic field (primarily motivated by the quantum Hall 
effect\cite{girvin}) to study the vortex states of FFLO superconductors. 
Using these techniques, we develop a very efficient 
method to determine the VLS implied by a given Ginsburg-Landau
free energy functional. As established by previous 
studies\cite{sr,krs,hb,hbbm,klein}, the VLS is highly
non-universal; we argue here that the VLS approaches a
universal limit near the tricritical point, in the sense that the structure 
near this point in the phase diagram depends on the LL index of the order parameter only.
%Allan
We determine
the VLS in this limit explicitly. We demonstrate that the order parameter spatial profile is
completely determined, up 
to a uniform translation, by its Landau level index $n$ and by the VLS, and 
derive an explicit expression for the order parameter spatial profile that can be used 
to determine $n$ from experimental data. 
If the LL index can be determined by comparing experimental and  
theoretical VLSs, the order parameter momentum in the absence of a field 
can also be determined.

We lay out the theoretical method we have developed in the following section, and 
present the VLSs suggested by its application in section III. These techniques may
be useful in addressing other aspects of high Landau level bosonic physics in 
FFLO superconductors or elsewhere. 
% Allan
In section IV
we summarize our results and make a comparison with related recent work. 
%New paragraph above. -- KUN 4/20

\section{Theoretical method}
 
To address FFLO VLS physics we follow a Ginzburg-Landau approach similar to that of Refs.\onlinecite{hb,hbbm}.
The free-energy density appropriate for the FFLO state,
has the form\cite{bk}:
\beqarr
&F&\propto|(-\nabla^2-q^2)\psi|^2+a|\psi|^2+b|\psi|^4+c|\psi|^2|\nabla\psi|^2\nonumber\\
&+&d[(\psi^*)^2(\nabla\psi)^2+\psi^2(\nabla\psi^*)^2]+e|\psi|^6,
\label{F}
\eeqarr
where $a,b,c,d,e$ and $q$ are parameters that depend on both temperature and
Zeeman splitting.  The free energy of Eq.(\ref{F}) is appropriate for an 
s-wave superconductor with completely isotropic gap and 
band dispersion, the case we will consider here. Anisotropy in either 
pairing interaction or single electron dispersion will lead to additional 
terms\cite{ay}. The fundamental difference between FFLO
and BCS superconductors is expressed by the first term in $F$ which describes the
{\em kinetic energy} of the order parameter; in an FFLO superconductor this 
term is minimized when the order parameter carries a finite wave vector 
(or momentum) $q$; in a BCS superconductor $q=0$.  Thus far we have only taken into account the 
Zeeman effect of the external magnetic field, which is appropriate for the case
of a two-dimensional superconductor with an external magnetic field parallel 
to the plane.  For a 3D superconductor or for a 2D superconductor with the field ${\bf B}$
tilted out of system plane, orbital coupling must be accounted for 
by performing a minimal substitution $\nabla\psi\rightarrow {\bf D}\psi =
(\nabla-2ie{\bf A}/c)\psi$ with 
$\nabla\times {\bf A}=B_{\perp} \hat{z}$.  It is the 2D case that we have in 
mind here because it allows the Zeeman field, which controls $q$ and the other
LG model parameters, and the field $B_{\perp}$ to be controlled independently.

To find the order parameter $\psi$ that minimizes the free energy of 
the system near a continuous phase transition, 
one first determines the Cooper pair Landau level index that minimizes 
the coefficient of the quadratic terms in Eq. (\ref{F}). 
The eigenvalues of ${\bf D}^2=(\nabla-2ie{\bf A}/c)^2$ are 
$-(2n+1)/\ell^2$, where $\ell=\sqrt{\hbar c/2eB}$ is the Cooper pair
magnetic length and $n=0, 1, 2,\cdots$ is the LL index. For a BCS superconductor the kinetic
energy is minimized by $n=0$, {\em i.e.}, $\psi$ is a lowest LL wave function.
For a FFLO superconductor, however, the kinetic energy
is minimized by the index $n$ that minimizes $|(2n+1)/\ell^2-q^2|$.  
The optimal value of $n$ will be large when $B_{\perp}$ is small.\cite{bul}

Because of the macroscopic degeneracy of a Landau level, the order parameter retains 
considerable freedom at $B_{\perp} \ne 0$ even when $n$ is fixed.  As we explain below,
this freedom is resolved by minimizing the remaining terms
of Eq. (\ref{F}) with a properly normalized wave function $\psi({\bf r})$; this 
corresponds to optimizing the vortex lattice.  The overall magnitude of the 
order parameter is determined by balancing quadratic and higher order terms in the GL energy
functional. For a BCS
superconductor that is described by an ordinary Ginsburg-Landau theory 
for which $n=0$ and in
which the $|\psi|^4$ term is the only addition term,
it is known that the optimal VLS is triangular.
In the following we use techniques developed in studies 
of the quantum Hall effect to derive expressions for and understand the relationships
between the higher order terms present in Eq. (\ref{F}). 

We assume that the VLS is that of a Bravais lattice so that there is one 
flux quantum per unit cell. The central quantity in our calculation is
the Fourier transform of $|\psi({\bf r})|^2$:
$\int{d^2{\bf r}}|\psi({\bf r})|^2e^{-i{\bf k}\cdot{\bf r}}$.
Since $|\psi({\bf r})|^2$ is periodic, this integral is zero unless 
${\bf k}={\bf G}$, where ${\bf G}$ is a reciprocal lattice vector of the 
vortex lattice. For ${\bf k}={\bf G}$, we define
\beq
\tilde{\Delta}_{\bf G}=\int_{u.c.}{d^2{\bf r}}
|\psi({\bf r})|^2e^{-i{\bf G}\cdot{\bf r}}=
\langle\psi|e^{-i{\bf G}\cdot{\bf r}}|\psi\rangle.
\label{fourier}
\eeq
Here $\int_{u.c.}$ stands for integration within a unit cell of the vortex 
lattice, and we
assume that $\psi({\bf r})$ is normalized within a unit cell so that
\beq
\int_{u.c.}{d^2{\bf r}}|\psi({\bf r})|^2=\langle\psi|\psi\rangle=1.
\eeq
To proceed, we introduce the guiding center coordinate\cite{haldane}
\beq
{\bf R}={\bf r}+\ell^2 \hat{z}\times(-i\nabla+(2e/c){\bf A}), 
\eeq
and separate 
${\bf r}$ into ${\bf r}={\bf R}+({\bf r}-{\bf R})$.
It is known\cite{girvin} that the guiding center ${\bf R}$ has no matrix element between states
in different LLs, while $({\bf r}-{\bf R})$ changes the LL index of a state without
changing its guiding center dependence; as a consequence components 
of ${\bf R}$ and $({\bf r}-{\bf R})$ commute with each other\cite{haldane}. 

% Allan: Small changes - separate paragraph to highlight. 
Two key observations simplify the calculations below. First $|\psi\rangle$ is an eigenket of
$e^{-i{\bf G}\cdot{\bf R}}$.  It follows that $\Delta_{\bf G}$ is the eigenvalue of 
this operator and, since the operator is unitary, must be a phase factor.  As we
% Final:  
discuss later the phase factor can be fixed by a convenient choice of spatial origin.
The operator $\exp(-i{\bf G}\cdot{\bf R})$ results in a translation
by ${\bf G}\ell^2 \times \hat{z}$ in real space, 
which is a direct lattice vector, and multiplication by a 
phase factor that is gauge but not position dependent.  The most straightforward way to prove 
this property is to construct $\psi({\bf r})$ explicitly in a convenient gauge (say the Landau 
gauge), show that it is true in this gauge,
and that an arbitrary gauge transformation (on $\psi({\bf r})$ and
$e^{-i{\bf G}\cdot{\bf R}}$ simultaneously)
does not change the result.  The second key observations follows from the fact that 
${\bf r}-{\bf R}$ does not act on guiding center coordinates: 
\beq
\langle\psi|e^{-i{\bf G}\cdot({\bf r}-{\bf R})}|\psi\rangle  = 
L_n(G^2\ell^2/2)e^{-G^2\ell^2/4}
\eeq
for {\em any} normalized $n$-th Landau level state. 
(Here $L_n(x)$ is the Laguerre polynomial.) 
 
It follows from the above properties that 
\beqarr
\tilde{\Delta}_{\bf G}&=&\langle\psi|e^{-i{\bf G}\cdot{\bf r}}|\psi\rangle
=\langle\psi|e^{-i{\bf G}\cdot{\bf R}}|\psi\rangle
\langle \psi|e^{-i{\bf G}\cdot({\bf r}-{\bf R})}|\psi\rangle
\nonumber\\
&=&\Delta_{\bf G}L_n(G^2\ell^2/2)e^{-G^2\ell^2/4},
\label{delta}
\eeqarr
with $|\Delta_{\bf G}|=1$. Eq. (\ref{delta}) allows for a very straightforward
calculation of the Abrikosov factor
\beqarr
\beta^{(n)}_A&=&A\int_{u.c.}{d^2{\bf r}}|\psi({\bf r})|^4
=\sum_{\bf G}|\tilde{\Delta}_{\bf G}|^2\nonumber\\
&=&\sum_{\bf G}[L_n(G^2\ell^2/2)]^2
e^{-G^2\ell^2/2},
\label{beta}
\eeqarr
where $A=2\pi\ell^2$ is the area of the unit cell which contains one flux 
quantum. In Eq. (\ref{beta}) the VLS is characterized by its reciprocal lattice vectors
${\bf G}$.
For BCS superconductors described by ordinary Ginsburg-Landau free energy
functional, the VLS is determined by minimizing $\beta^{(0)}_A$;\cite{saint}
it is known that in the lowest LL ($n=0$), $\beta^{(0)}_A$ is 
minimized by the triangular lattice with $\beta^{(0)}_A=1.1596$,\cite{saint} a 
result easily reproduced using Eq. (\ref{beta}), 
%Added stuff below -- KUN 7/15
with the choice 
${\bf G}=m{\bf g}_1+n{\bf g}_2$, 
where ${\bf g}_1=(\sqrt{4\pi/\sqrt{3}}/\ell)\hat{x}$ 
and ${\bf g}_2=(\sqrt{\pi/\sqrt{3}}/\ell)\hat{x}
+(\sqrt{\sqrt{3}\pi}/\ell)\hat{y}$ are the
basis vectors of the (reciprocal) triangular lattice.
For the square lattice, we have ${\bf g}_1=(\sqrt{2\pi}/\ell)\hat{x}$
and ${\bf g}_2=(\sqrt{2\pi}/\ell)\hat{y}$, which yields
$\beta^{(0)}_A=1.1803$, also in agreement with previously known results.\cite{saint}

%Allan - small changes 
To determine the VLS of a FFLO superconductor one needs to take into 
account the fact that the LL index $n$ is not necessarily zero and
the presence of contributions from terms other than $|\psi({\bf r})|^4$ in the free energy
functional (\ref{F}). These can be done straightforwardly using
extensions of Eq. (\ref{beta}) that also follow from the separation of 
guiding-center dependence in the appropriate operators. 
For example the contribution 
from the $|\psi|^6$ term is parametrized by 
\beqarr
\gamma^{(n)}_A&=&A^2\int_{u.c.}{d^2{\bf r}}|\psi({\bf r})|^6=\sum_{{\bf G}_1{\bf G}_2}
\tilde{\Delta}_{{\bf G}_1}\tilde{\Delta}_{{\bf G}_2}
\tilde{\Delta}_{-{\bf G}_1-{\bf G}_2}\nonumber\\
&=&\sum_{{\bf G}_1{\bf G}_2}
\Delta_{{\bf G}_1}\Delta_{{\bf G}_2}
\Delta_{-{\bf G}_1-{\bf G}_2}
e^{-(G_1^2+G_2^2+{\bf G}_1\cdot{\bf G}_2)\ell^2/2}\nonumber\\
&\times& L_n(G_1^2\ell^2/2)L_n(G_2^2\ell^2/2)L_n(|{\bf G}_1+{\bf G}_2|^2\ell^2/2)\nonumber\\
&=&\sum_{{\bf G}_1{\bf G}_2}
\cos[\hat{z}\cdot({\bf G}_1\times{\bf G}_2)\ell^2/2]
e^{-(G_1^2+G_2^2+{\bf G}_1\cdot{\bf G}_2)\ell^2/2}\nonumber\\
&\times& L_n(G_1^2\ell^2/2)L_n(G_2^2\ell^2/2)L_n(|{\bf G}_1+{\bf G}_2|^2\ell^2/2),
\label{gamma}
\eeqarr
where we used the property 
\beq
e^{-i({\bf G}_1+{\bf G}_2)\cdot{\bf R}}=e^{-i{\bf G}_1\cdot{\bf R}}
e^{-i{\bf G}_2\cdot{\bf R}}e^{i\hat{z}\cdot({\bf G}_1\times{\bf G}_2)\ell^2/2},
\eeq
which follows from the commutator between different components of ${\bf R}$\cite{haldane}:
$[R_x, R_y]=i\ell^2$.  It follows that
\beq
\Delta_{-{\bf G}_1-{\bf G}_2}=\Delta^{-1}_{{\bf G}_1}\Delta^{-1}_{{\bf G}_2}
e^{i\hat{z}\cdot({\bf G}_1\times{\bf G}_2)\ell^2/2}.
\eeq
%New equation added above. -- KUN 4/20

Similarly, using the fact that $\psi$ is an eigenfunction of the operator 
$(\nabla-2ie{\bf A}/c)^2$,
the contributions from the $c$ and $d$ terms in Eq. (\ref{F}) can be expressed
as combinations of $\beta^{(n)}_A$ and\cite{hb}  
\beq
I^{(n)}_{22}=\int_{u.c.}{d^2{\bf r}}|\psi({\bf r})|^2|{\bf D}\psi({\bf r})|^2.
\eeq
To evaluate $I^{(n)}_{22}$, we introduce
\beqarr
\lambda_{\bf G}&=&\int_{u.c.}{d^2{\bf r}}
|{\bf D}\psi({\bf r})|^2e^{-i{\bf G}\cdot{\bf r}}\nonumber\\
&=&-\int_{u.c.}{d^2{\bf r}}
\psi^*{\bf D}[{\bf D}\psi({\bf r})e^{-i{\bf G}\cdot{\bf r}}]\nonumber\\
&=&-\int_{u.c.}{d^2{\bf r}}
\psi^*[{\bf D}^2\psi({\bf r})-i({\bf G}\cdot{\bf D})\psi({\bf r})]e^{-i{\bf G}\cdot{\bf r}}.
\label{lambda}
\eeqarr
It is easy to show using the fact that $\psi$ is an eigen wavefunction of
${\bf D}^2$ that the second term in the final line of Eq. (\ref{lambda}) 
does not contribute to the integral. We thus find 
\beq
\lambda_{\bf G}=(2n+1)\tilde{\Delta}_{\bf G}/\ell^2,
\eeq
and
\beq
I^{(n)}_{22}=\sum_{\bf G}\tilde{\Delta}_{\bf G}\lambda_{-{\bf G}}
=(2n+1)\beta^{(n)}_A/\ell^2.
\eeq
We have thus found that $I^{(n)}_{22}$ is proportional to $\beta^{(n)}_A$; 
in particular, they have exactly the same dependence on the VLS.
This fact simplifies the determination of the optimal VLS significantly.

\begin{figure*}
\includegraphics[width = 1\textwidth]{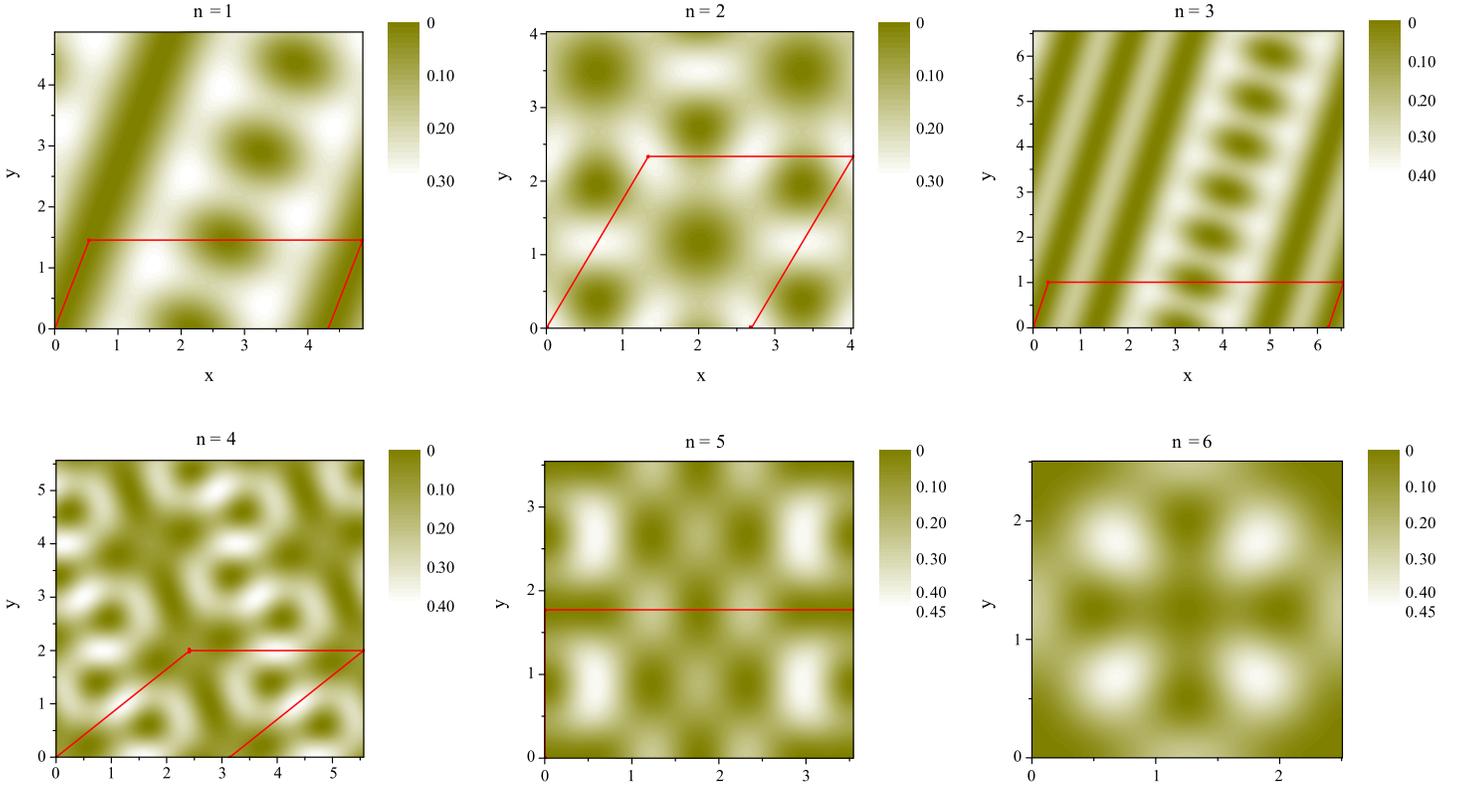}
%\epsfxsize=6in
%\centerline{\epsffile{fig.eps} }
\caption{(Color online)
Plots of the square of order parameter magnitude that minimizes
$\beta_A^{(n)}$, for Landau levels $0 < n \le 6$.
The lines (when present) indicate the region of a unit cell of the vortex
lattice. Lengths are in unit of the magnetic length $\ell$.
}
\label{figure}
\end{figure*}

\section{Results}

In tables I and II we have listed the VLS that minimizes $\beta^{(n)}_A$ 
and $\gamma^{(n)}_A$ respectively, for the order parameter in 
the $n$th Landau level. Fig. \ref{figure} contains plots for small 
Landau level indices of the square of
the order parameter magnitude  
for the vortex lattices that minimize $\beta^{(n)}_A$.
%Figure added here. KUN 4/20
For $n=0$ we find that
that both $\beta_A$ and $\gamma_A$ are minimized by the triangular lattice,
in agreement with previously known results\cite{saint,hb}. 
% Final: Do we have room for a figure here? 
% Allan 
Evidently the situation changes drastically for $n \ne 0$.
The competition between different potential VLS's  can be  
understood in terms of the simple reciprocal lattice expression of $\beta^{(n)}_A$, 
Eq. (\ref{beta}). For $n=0$, the Laguerre polynomial is $L_{n=0}(x) \equiv 1$, so it 
is clear that $\beta_A$ is minimized by placing the reciprocal lattice vectors
(${\bf G}$'s) as far away from the origin as possible given the 
fixed unit cell area, which is what a 
triangular lattice does. For
$n\ge 1$, on the other hand, $L_{n}(x)$ is a polynomial with nodes whose positions depend on 
$n$; one may thus lower 
$\beta^{(n)}_A$ by, for example, choosing ${\bf G}$'s such that 
$L_n(G^2\ell^2/2)$ is close to zero for some reciprocal lattice vectors, leading to very non-standard VLSs 
under most circumstances.  Similar considerations apply to 
the optimal VLS that minimizes $\gamma^{(n)}_A$.
Overall, the optimal values of both $\beta^{(n)}_A$
and $\gamma^{(n)}_A$ increase with $n$, reflecting the fact that the order
parameter wave function becomes increasingly oscillatory (and therefore 
less uniform) as $n$ increases.  For $n=1$ and $3$ VLS's are characterized by 
order parameter concentration in widely spaced stripe-like regions  
with rapid but weak longitudinal modulations.  The stripes are characterized by lines with 
definite guiding centers and odd parity perpendicular quantum wave functions, yielding
a series of lines along which the order parameter is small.  For even $n$ and for larger $n$ the optimal 
VLS's tend to be more close-packed, with lattices that are close to
square occurring frequently.

\begin{table}
\begin{ruledtabular}
\begin{tabular}{c c c c c c c c}
LL index  & 0 & 1 & 2 & 3 & 4 & 5 & 6 \\
\hline
aspect ratio & 1 & 0.36 & 1 &     0.17    &     1    &     0.50    &  1 \\
angle $\theta$ & $\pi/3$ & 1.21 & 1.05 & 1.26 & 0.69  & $\pi/2$ & $\pi/2$  \\
$\beta^{(n)}_A$ & 1.16 & 1.32 & 1.20 & 1.43 & 1.38 & 1.42  & 1.44
\\
\end{tabular}
\end{ruledtabular}
\caption{The optimal vortex lattice structure that minimizes $\beta^{(n)}_A$
in different Landau levels. The vortex lattice structure is
parametrized by the aspect ratio between the two basis vectors, and the angle
between them.}
\label{tab:beta}
\end{table}

\begin{table}
\begin{ruledtabular}
\begin{tabular}{c c c c c c c c}
LL index  & 0 & 1 & 2 & 3 & 4 & 5 & 6 \\
\hline
aspect ratio & 1 & 0.39 & 1 &  0.16    &  1  & 1  &  1   \\
angle $\theta$ & $\pi/3$ & 1.53 & 1.04 &     1.10    & 0.68  & $\pi/3$ & $\pi/2$ \\
$\gamma^{(n)}_A$ & 1.42 & 1.89 & 1.57 & 2.29  & 2.19 & 2.31  & 2.47 \\
\end{tabular}
\end{ruledtabular}
\caption{The optimal vortex lattice structure that minimizes $\gamma^{(n)}_A$
in different Landau levels. The vortex lattice structure is
parametrized by the aspect ratio between the two basis vectors, and the angle
between them.}
\label{tab:gamma}
\end{table}

Since both the quartic terms ($b, c$ and $d$ terms in Eq. (\ref{F})) and the
sixth order term (the $e$ term) contribute to the total free energy, and
$\beta^{(n)}_A$ and $\gamma^{(n)}_A$ depend on the lattice structure in different ways for
$n\ge 1$, the optimal VLS will depends on an interplay between the two, and is therefore {\em non-universal};
this is very different from the Abrikosov lattice of the BCS superconductor,
which is always triangular. In the following we discuss two limiting cases
in which the optimal VLS approaches a universal limit, in the sense that it 
depends on the Landau level index $n$ only. 

(i) It is known\cite{bk} that at 
the tricritical point which occurs for $T\approx 0.56 T_c$ (the maximum temperature 
at which the FFLO state is stable) 
{\em both} the order parameter wave vector $q$ and the 
coefficient of the quartic term $b$ vanish. Near this tricritical point,
the contributions of the quartic terms to the free energy vanishes 
(the terms proportional to $c$ and $d$ both involve gradients are proportional to $q^2$).
The optimal VLS is therefore determined by minimizing $\gamma^{(n)}_A$ alone 
and
is given by the entries listed in table II. 

(ii) Along the 
$H_{c2}$ line, the transition can be either first order or second order. 
The transition is second order when the combined contribution to the free
energy from the quartic terms is positive, in which case the contribution from
sixth order term is negligible near the phase boundary, due to the smallness
of the order paramter. As shown above, the contribution to the free energy
from all the quartic terms is proportional to $\beta^{(n)}_A$ with {\em lattice
structure independent prefactors}.
Minimizing the free energy is therefore equivalent to
minimizing $\beta^{(n)}_A$, resulting in an optimal VLS that depends on
$n$ only and given in this case by the entries listed in table I. 
There is, however, an important caveat in this
case since the conclusion is reached based on the free energy Eq. (\ref{F}), in 
which quartic terms that involve more gradients are neglected; these terms
are in principle comparable to the terms kept, because the order parameter
wavevector $q$ is finite, and it may not be possible to express their 
contributions in terms of $\beta^{(n)}_A$ alone. These terms do not affect 
case (i) however because $q$ approaches zero there.

Once the VLS is determined, one can obtain the detailed spatial dependence of
the order parameter in the $n$th LL
by counter-Fourier transform Eq. (\ref{fourier}):
\beqarr
|\psi({\bf r})|^2&=&{1\over 2\pi\ell^2}\sum_{\bf G}\tilde{\Delta}_{\bf G}
e^{-i{\bf G}\cdot{\bf r}}\nonumber\\
&=&{1\over 2\pi\ell^2}\sum_{\bf G}\Delta_{\bf G}L_n(G^2\ell^2/2)e^{-G^2\ell^2/4}
e^{i{\bf G}\cdot{\bf r}}.
\eeqarr
By parameterizing the reciprocal lattice vector ${\bf G}=m_1{\bf g}_1+m_2{\bf g}_2$ 
where ${\bf g}_1$ and ${\bf g}_2$ are the two basis vectors of the 
reciprocal lattice, while $m_1$ and $m_2$ are two integers, we find
\beqarr
&&\Delta_{\bf G}=\langle\psi|e^{-i(m_1{\bf g}_1+m_2{\bf g}_2)\cdot{\bf R}}|\psi\rangle
%\nonumber\\
=e^{im_1m_2\hat{z}\cdot({\bf g}_1\times{\bf g}_2)\ell^2/2}\nonumber\\
&\times&\langle\psi|e^{-im_1{\bf g}_1\cdot{\bf R}}e^{-im_2{\bf g}_2\cdot{\bf R}}|\psi\rangle
%\nonumber\\
=(-1)^{m_1m_2}e^{i(m_1\theta_1+m_2\theta_2)},
\eeqarr
where $e^{i\theta_1}$ and $e^{i\theta_2}$ are the eigen values of 
$e^{-i{\bf g}_1\cdot{\bf R}}$ and $e^{-i{\bf g}_2\cdot{\bf R}}$
respectively; 
% Final
we may choose $\theta_1=\theta_2=0$ by selecting the spatial origin in the unit cell.
We thus obtain
\beqarr
|\psi({\bf r})|^2&=&{1\over 2\pi\ell^2}\sum_{m_1m_2} (-1)^{m_1m_2}
L_n(|m_1{\bf g}_1+m_2{\bf g}_2|^2\ell^2/2)
\nonumber\\
&\times& e^{-|m_1{\bf g}_1+m_2{\bf g}_2|^2\ell^2/4}
\cos[(m_1{\bf g}_1+m_2{\bf g}_2)\cdot
{\bf r}].
\label{gap}
\eeqarr

Experimentally one can determine the VLS (and thus ${\bf g}_1$ and ${\bf g}_2$),
by neutron or muon scattering for example, while the magnitude of the
local order parameter (or gap) may be measured by STM.  
By comparing the measured
results with Eq. (\ref{gap}) one may be able to determine the LL index $n$.
This allows for an estimate of the order parameter wavevector in the absence of
the orbital magnetic field: $q\approx\sqrt{2n+1}/\ell$, which is an alternative
to measuring $q$ using the Josephson effect\cite{ya}.
%Caveats pointed out below. KUN 4/20
It is important however to keep the following caveats in mind when making this
comparison. (i) We have assumed here that the order parameter wave function 
is in a given Landau level exclusively. This is true only when one is very 
close to the portions of $H_{c2}$ line where the transition is second order,
or to the tricritical point mentioned above. Away from these regions mixing 
of different Landau levels becomes increasingly important, and this can 
affect our results significantly. 
(ii) Our analysis is based on a Ginsburg-Landau free energy that is completely isotropic. Real systems often have
some anisotropy,  due to either the pairing interaction (as in a d-wave
superconductor) or single-electron dispersion (band structure effect). It
should be possible however to generalize our methods to anisotropic cases.

\section{Summary}
In this work we have presented and applied an efficient method to determine the vortex 
lattice structure (VLS) of an FFLO superconductor, based on an appropriate Ginsburg-Landau 
free energy functional. In our approach all calculations are 
performed in a completely gauge independent manner. We have shown that the
VLS is universal in certain limiting cases, in the sense that it depends on the
Landau level index of the order parameter only.
% Allan:  This is a good point - trying to `thump' a little harder. 
% Also placed this remark in introduction and abstract.  
Independent of the VLS, which is not universal in the general case,
our results enable an experimental determination of the Landau level
index which follows from the constraints on spatial structure that exist when only 
guiding center degrees of freedom are available.  Given the Landau level index,
one can estimate the order parameter wavevector in the absence of orbital magnetic field.

%new paragraph below. -- KUN 4/20
Fundamentally, our approach is similar to
that used in Refs. \onlinecite{hb,hbbm} in that it is 
based on minimizing an appropriate Ginsburg-Landau
free energy functional. The efficiency of the new calculational technique,
however, allows one to determine the vortex lattice structure for a wide range
of parameters, instead of only the specific cases studied earlier. 
Our methods can be applied to any problem in which bosons 
occupy high Landau levels.  In addition to vortex states of FFLO 
superconductors, another possibility is bosonic cold atoms in a 
rotating trap with an optical lattice potential that has been 
engineered to have the minimum band energy away from the zone 
center. 

In a recent paper,\cite{klein} Klein studied the VLS of FFLO superconductors
starting directly from a weak-coupling BCS model 
%(......) added below. KUN 7/15
(which is a generalization of earlier work by Eilenberger\cite{eilenberger}),
instead of using a Ginsburg-Landau
free energy functional. In his work the VLS was determined solely from the
terms that are quartic in the order parameter
in the free energy, while as we see here terms beyond
quartic can be crucial in certain regimes. On the other hand it appears that
he was able to evaluate the quartic terms {\em without} using a
gradient expansion, 
which appears to be an improvement on the usual Ginsburg-Landau approach;
this might be important under some circumstances. 
We note that we find the dependence of VLS on the Landau level index $n$
appears to be quite complicated, without any obvious
indication of the simplification in the
limit $n\rightarrow\infty$, suggested in Ref. \onlinecite{klein}.

%Comments added above. -- KUN 4/20

\acknowledgements
We thank Q. Cui for technical assistance.
KY was supported by NSF grant No. DMR-0225698, and by the Research
Corporation. AHM was supported by NSF grant No. DMR-0115947 and by the 
Welch Foundation.

\end{document}